\documentclass{PoS}
\newcommand{\SO}{\mathrm{SO}}
\newcommand{\dd}{{\rm{d}}}
\newcommand{\tr}{{\rm Tr\,}}

\title{The numerical approach to quantum field theory in a non-commutative space}

\ShortTitle{The numerical approach to quantum field theory in a non-commutative space}

\author{\speaker{Marco~Panero}\\
        Department of Physics, University of Turin and INFN, Turin\\
        Via Pietro Giuria 1, I-10125 Turin, Italy\\
        E-mail: \email{marco.panero@unito.it}}

\abstract{Numerical simulation is an important non-perturbative tool to study quantum field theories defined in non-commutative spaces. In this contribution, a selection of results from Monte~Carlo calculations for non-commutative models is presented, and their implications are reviewed. In addition, we also discuss how related numerical techniques have been recently applied in computer simulations of dimensionally reduced supersymmetric theories.}

\FullConference{Proceedings of the Corfu Summer Institute 2015 "School and Workshops on Elementary Particle Physics and Gravity"\\
1-27 September 2015\\
Corfu, Greece}

\begin{document}

\section{Introduction}
\label{sec:introduction}

The idea that the geometry of spacetime becomes non-commutative at the Planck scale is a central feature of quantum gravity theories~\cite{Doplicher:1994tu}: it arises naturally in a string theory context~\cite{Seiberg:1999vs} and entails far-reaching consequences for quantum field theory (QFT), including a variety of surprising phenomenological implications~\cite{Douglas:2001ba, Szabo:2001kg}; a well-known example of the latter is the mixing of ultraviolet and infrared (UV/IR) degrees of freedom~\cite{Minwalla:1999px}.

Typically, analytical studies of QFT in non-commutative spaces (NC~QFT) involve some form of approximation---be it a truncated perturbative expansion, or some large-$N$ limit, et c. In order to test the robustness of results obtained under these approximations, it is desirable to check them against some other \emph{ab initio} formulation of the theory. For a generic NC~QFT, a possible way to do this is by numerical evaluation of correlation functions, in a Feynman path-integral formulation of the theory~\cite{Feynman:1948ur}. During the past fifteen years, this approach has been successfully pursued in several works, in which various types of field theories in non-commutative spaces of different dimensions have been mapped to appropriately defined finite-dimensional matrix models, and investigated by Monte~Carlo integration. More recently, the numerical technology developed in these works has also found applications in computer simulations of dimensionally reduced supersymmetric theories, paving the way to the non-perturbative study of many open theoretical issues---including, in particular, problems related to the gauge/gravity duality~\cite{Maldacena:1997re, Gubser:1998bc, Witten:1998qj}.

\section{Implementation}
\label{sec:implementation}

In the Feynman path-integral formalism, expectation values in a generic QFT (defined in ordinary, commutative Minkowski spacetime) are given by
\begin{equation}
\label{Minkowskian_Feynman_path_integral}
\langle \mathcal{O}^{\mbox{\tiny{(M)}}}_1 \dots \mathcal{O}^{\mbox{\tiny{(M)}}}_n \rangle = \frac{\int \mathcal{D} \phi \; \mathcal{O}^{\mbox{\tiny{(M)}}}_1 \dots \mathcal{O}^{\mbox{\tiny{(M)}}}_n \exp \left( i S^{\mbox{\tiny{(M)}}} \right)}{\int \mathcal{D} \phi \; \exp \left( i S^{\mbox{\tiny{(M)}}} \right)}.
\end{equation}
In a Euclidean formulation, the previous expression is replaced by
\begin{equation}
\label{Euclidean_Feynman_path_integral}
\langle \mathcal{O}_1 \dots \mathcal{O}_n \rangle = \frac{\int \mathcal{D} \phi \;\mathcal{O}_1 \dots \mathcal{O}_n \exp \left( - S \right)}{\int \mathcal{D} \phi \; \exp \left( -S \right)},
\end{equation}
where $S$ denotes the Euclidean action, a functional of the fields $\phi$. The similarity between eq.~(\ref{Euclidean_Feynman_path_integral}) and the expression for correlation functions in a statistical system suggests that the functional integrals appearing on the r.h.s. of eq.~(\ref{Euclidean_Feynman_path_integral}) could be evaluated using techniques analogous to the computational tools of statistical mechanics, including low- or high-temperature expansions, or numerical integration by Monte~Carlo methods.

The latter approach requires the definition of a measure for the fields, that is both (\emph{i}) mathematically well-defined, and (\emph{ii}) suitable for numerical calculations. In practice, this means that the fields in the original theory have to be traded for a finite number of degrees of freedom, often defined in terms of matrix variables. This procedure is straightforward for bosonic fields, which are represented by ordinary, commuting $c$-numbers in the Feynman path-integral formalism. To enforce the correct statistics for fermionic fields, on the other hand, they have to be represented by Gra{\ss}mann numbers; even though a direct computer evaluation of ``ensemble averages'' of products of Gra{\ss}mann variables is not possible, the integration over these quantities can be carried out exactly for a Euclidean action which depends bilinearly on the fermionic fields. This fermionic Gau{\ss}ian integral results into the determinant of the large, but finite-dimensional, matrix (a discretized version of the Dirac operator in the original theory), while fermionic operators appearing in a generic observable $\mathcal{O}$ are associated with  elements of the inverse of such matrix~\cite{Matthews:1954zg, Matthews:1955zi}. Note that, in general, both the determinant and the inverse matrix elements are highly non-local, yet completely well-defined functions of the bosonic degrees of freedom to which the fermionic fields are coupled. Thus, as long as the purely bosonic contribution to the Euclidean action is a real quantity bounded from below, and the fermionic matrix determinant is positive (a requirement that holds under certain conditions), each possible set of values of the bosonic degrees of freedom (a \emph{configuration}) is associated with a properly normalized, real positive Boltzmann weight, and the ``discretized'' version of eq.~(\ref{Euclidean_Feynman_path_integral}) describes a completely well-defined statistical-mechanics model, with a finite number of degrees of freedom. Typically, the  Boltzmann weight is a strongly peaked function in configuration space, so the high-dimensional integrals defining the correlation functions can be evaluated numerically, by Monte~Carlo sampling. Note that the ``discretization'' of the original theory to a finite number of degrees of freedom introduces an intrinsic cutoff; the original theory is then recovered in the limit in which the ``matrix size'' (or, more generally, the number of degrees of freedom of the discretized version of the system) is taken to infinity. More precisely, in this limit the original theory arises as \emph{a good low-energy effective description} of the discretized model---where ``low-energy'' means ``at scales that are well-separated from the cutoff scale''.

Let us now see how this can be done for path integrals in NC~QFT. In general, the details of the regularization in terms of finite matrices depend on the model under consideration: the simplest examples are provided by NC scalar field theory in two dimensions. If one defines this theory on a fuzzy sphere~\cite{Madore:1991bw}, then the scalar field can be directly mapped to a Hermitian matrix $\Phi$ of finite size $N$, and the Euclidean action takes the form
\begin{equation}
\label{fuzzy_sphere_action}
S = \frac{4\pi }{N} \tr \left( \Phi \left[ L_i , \left[ L_i, \Phi \right] \right] + r R^2 \Phi^2 + \lambda R^2 \Phi^4 \right),
\end{equation}
where $R$ is the sphere radius, $L_i$ (with $i=1$, $2$, $3$) denotes a generator of the $su(2)$ algebra, in the representation of spin $j=(N-1)/2$, while $r$ and $\lambda$ are the rescaled coefficients of the quadratic and quartic terms in the potential, respectively. The computation of correlation functions is then expressed in terms of integrals over the entries of the $\Phi$ matrix
\begin{equation}
\langle \mathcal{O}_1 \dots \mathcal{O}_n \rangle = \frac{\int \prod_{i,j=1}^N \dd \Phi_{ij} \;\mathcal{O}_1 \dots \mathcal{O}_n e^{-S}}{\int \prod_{i,j=1}^N \dd \Phi_{ij} e^{-S}}.
\end{equation}
This type of calculation has been carried out by Monte~Carlo methods in a number of works~\cite{Martin:2004un, Panero:2006bx, Panero:2006cs, GarciaFlores:2009hf, Digal:2011hf} (including a very recent study of the entanglement entropy~\cite{Okuno:2015kuc}).

A different way to map NC scalar field theory in two dimensions to a finite-dimensional matrix model was introduced in ref.~\cite{Ambjorn:1999ts}, following an approach which is related to large-$N$ volume reduction in the twisted Eguchi-Kawai model~\cite{GonzalezArroyo:1982ub, GonzalezArroyo:2010ss} (see also ref.~\cite[subsection~4.7]{Lucini:2012gg} for a detailed discussion). In this case, the Euclidean action of the discretized model reads
\begin{equation}
S = \tr \left[ \frac{1}{2} \sum_{\mu = 1}^{2} \left( \Gamma_{\mu} \Phi \Gamma_{\mu}^{\dagger} - \Phi \right)^{2} + \frac{\bar m^{2}}{2} \Phi^{2} + \frac{\bar \lambda}{4} \Phi^{4} \right],
\end{equation}
where the kinetic term involves the ``twist eaters'' $\Gamma_{\mu}$, which are $N \times N$ unitary matrices satisfying the 't~Hooft-Weyl algebra, $\Gamma_{\mu} \Gamma_{\nu} = z_{\nu\mu} \Gamma_{\nu} \Gamma_{\mu}$, with $z_{12} = z_{21}^{\star} = -e^{i \pi /N}$. Also this formulation of NC scalar field theory in two dimensions has been studied numerically in various works~\cite{Ambjorn:2002nj, Bietenholz:2002ev, Mejia-Diaz:2014lza}.

Before moving on to a selection of results from numerical studies of NC~QFT, we conclude this section comparing these types of discretization with the regularization of QFT on a spacetime grid that is performed in lattice field theory~\cite{Wilson:1974sk}. As it was suggested in the literature~\cite{Grosse:1995pr, Balachandran:2005ew}, the discretization of NC~QFT in terms of matrix models could provide an interesting and viable alternative to computations based on the lattice regularization (even for theories defined in ordinary, commutative spaces, if the non-commutativity parameter can be made sufficiently small, so that results can be eventually extrapolated to the commutative limit): it is therefore important to assess similarities and differences between the two approaches.
\begin{itemize}
\item Perhaps the most striking difference with the lattice regularization is that matrix-model discretizations of NC field theories can often be formulated in a way that preserves the spacetime symmetries exactly \emph{at every value of the cutoff}. A very clear example is provided by the action for the scalar theory on the fuzzy sphere defined in eq.~(\ref{fuzzy_sphere_action}), which is explicitly invariant under continuum $\SO(3)$ rotations at every value of $N$. By contrast, the lattice breaks the group of Euclidean rotations down to a discrete subgroup---typically, the dihedral (in $D=2$ spacetime dimensions), octahedral (in $D=3$) or hyperoctahedral (in $D=4$) group, if the lattice is a regular square, cubic or hypercubic grid. Similarly, the lattice also breaks the group of continuum translations down to the subgroup of translations by integer multiples of the lattice spacing $a$ in each direction. This implies that, on the lattice, the spacetime symmetries of the continuum theory are explicitly broken by discretization artifacts, and get restored \emph{only in the continuum limit} $a \to 0$.
\item The treatment of fermionic fields (especially as it concerns chirality and anomalies) is somewhat simpler in matrix-model formulations of NC field theories~\cite{Grosse:1994ed, Grosse:1995jt, Balachandran:1999qu, Balachandran:2003ay, Dolan:2007uf}, whereas all ultralocal, chirally symmetric lattice formulations of the Dirac operator lead to unphysical doubler modes~\cite{Nielsen:1980rz, Nielsen:1981xu}, which can be removed at the cost of sacrificing either exact chiral symmetry at finite lattice spacing~\cite{Wilson:1975hf} or the ultralocality of the Dirac operator~\cite{Kaplan:1992bt, Narayanan:1994gw, Neuberger:1997fp}.
\item Also the construction of supersymmetric models is usually simpler~\cite{Nishimura:2009xm, Joseph:2015xwa}, while a ``direct'' implementation of supersymmetry on the lattice requires fine-tuning~\cite{Bergner:2013nwa}, and more sophisticated formulations of lattice supersymmetry, which involve twisted formulations or orbifold constructions, are actually closer to matrix-model formulations of NC field theory~\cite{Catterall:2009it}.
\item The typical computational costs of present NC field theory simulations are not prohibitive. Most state-of-the-art lattice QCD calculations, instead, require supercomputing resources. 
\item The reason for the latter difference is not in intrinsic limitations of the lattice formulation, but rather in the fact that the primary focus of numerical studies of NC models is on \emph{qualitative} features of theories beyond the Standard Model, typically at energy scales far from the reach of present experiments, whereas lattice calculations are now in a precision era, and aim at accurate \emph{quantitative} results for QCD phenomenology~\cite{Kronfeld:2012uk}---while topics beyond QCD (e.g. large-$N$ gauge theories ~\cite{Panero:2012qx, Lucini:2013qja}, strongly coupled gauge theories for dynamical electro-weak symmetry breaking~\cite{DelDebbio:2010zz, Giedt:2012it}, et c.) are covered in a sizable, but much smaller, fraction of the lattice literature.
\end{itemize} 

\section{Examples of results}
\label{sec:results}

In this section, we review a (limited) selection of results from recent numerical studies of NC models. We start from the results obtained in simulations of scalar field theories in subsection~\ref{subsec:results_scalar}, before moving to those for non-supersymmetric gauge theories in subsection~\ref{subsec:results_gauge}, and finally discussing those for supersymmetric models in subsection~\ref{subsec:results_supersymmetry}.

\subsection{Results for NC scalar field theory}
\label{subsec:results_scalar}

\begin{figure}
\begin{center}
\includegraphics*[width=0.45\textwidth]{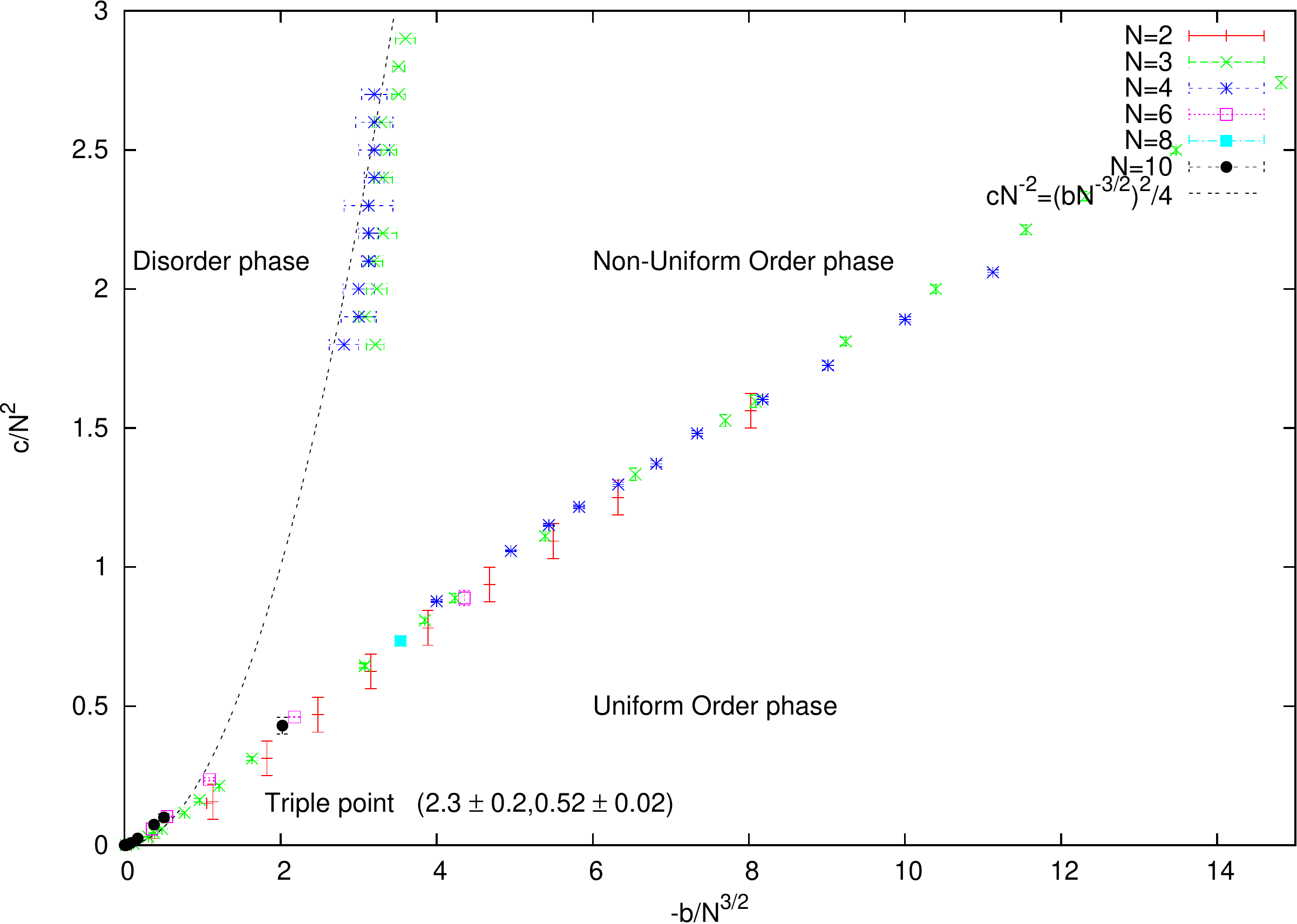} \hfill \includegraphics*[width=0.45\textwidth]{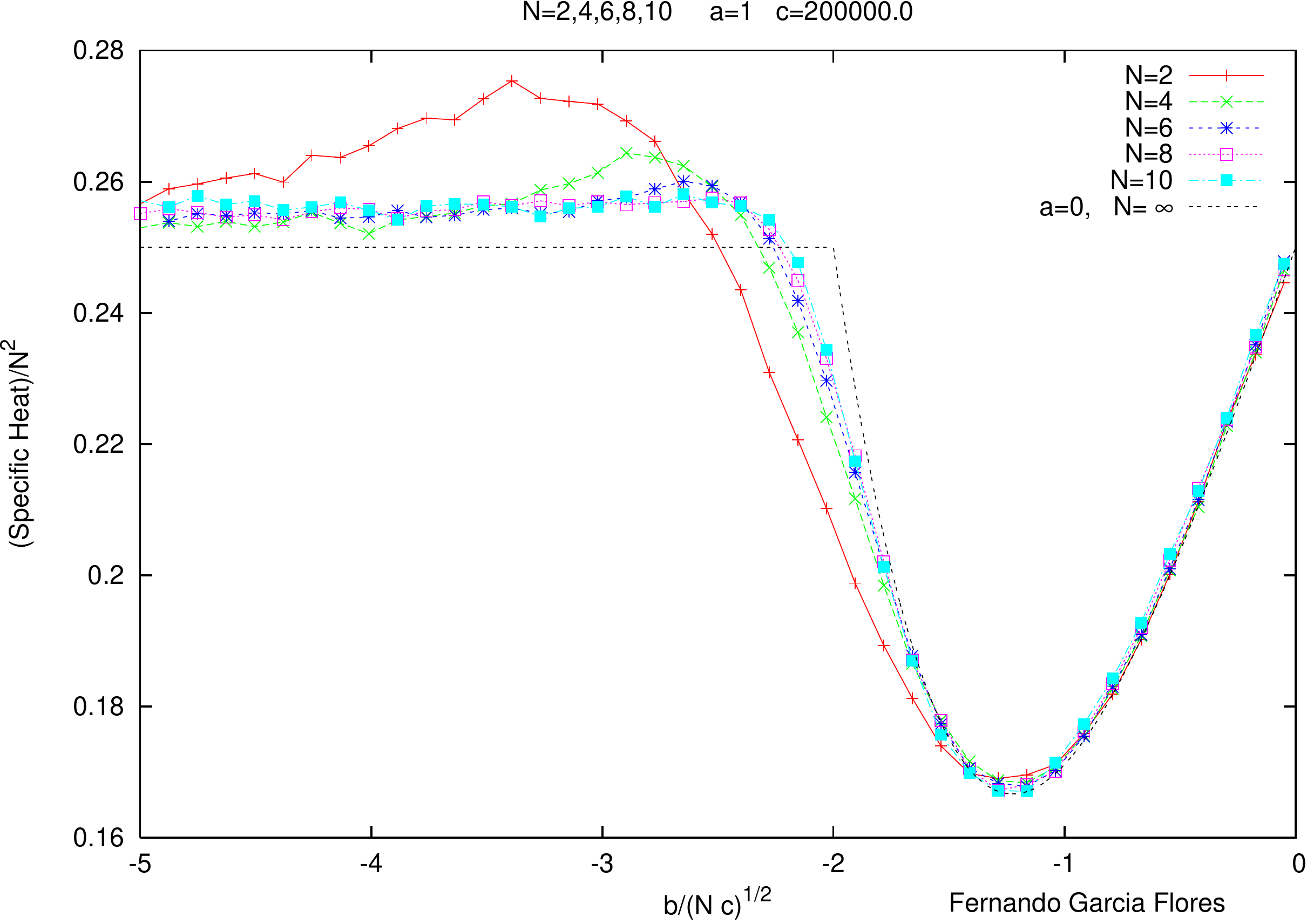}
\caption{\label{fig:phi4_phase_diagram_specific_heat} The phase structure (left-hand-side panel) and the scaling of the specific heat at the disorder/non-uniform order transition with the matrix size $N$, and its comparison with analytical expectations in the large-$N$ limit (right-hand-side panel) of NC $\phi^4$ theory in two dimensions, from the simulations on a fuzzy sphere reported in ref.~\cite{GarciaFlores:2009hf}.}
\end{center}
\end{figure}

Several works~\cite{Martin:2004un, Panero:2006bx, Panero:2006cs, GarciaFlores:2009hf, Mejia-Diaz:2014lza, Bietenholz:2014sza} studied the \emph{phase structure} of NC $\phi^4$ theory in two dimensions, finding numerical evidence for a \emph{striped phase}~\cite{Gubser:2000cd} characterized by non-uniform order. As an example, fig.~\ref{fig:phi4_phase_diagram_specific_heat} shows the phases and the specific heat obtained in ref.~\cite{GarciaFlores:2009hf}.

\begin{figure}
\begin{center}
\includegraphics*[width=0.48\textwidth]{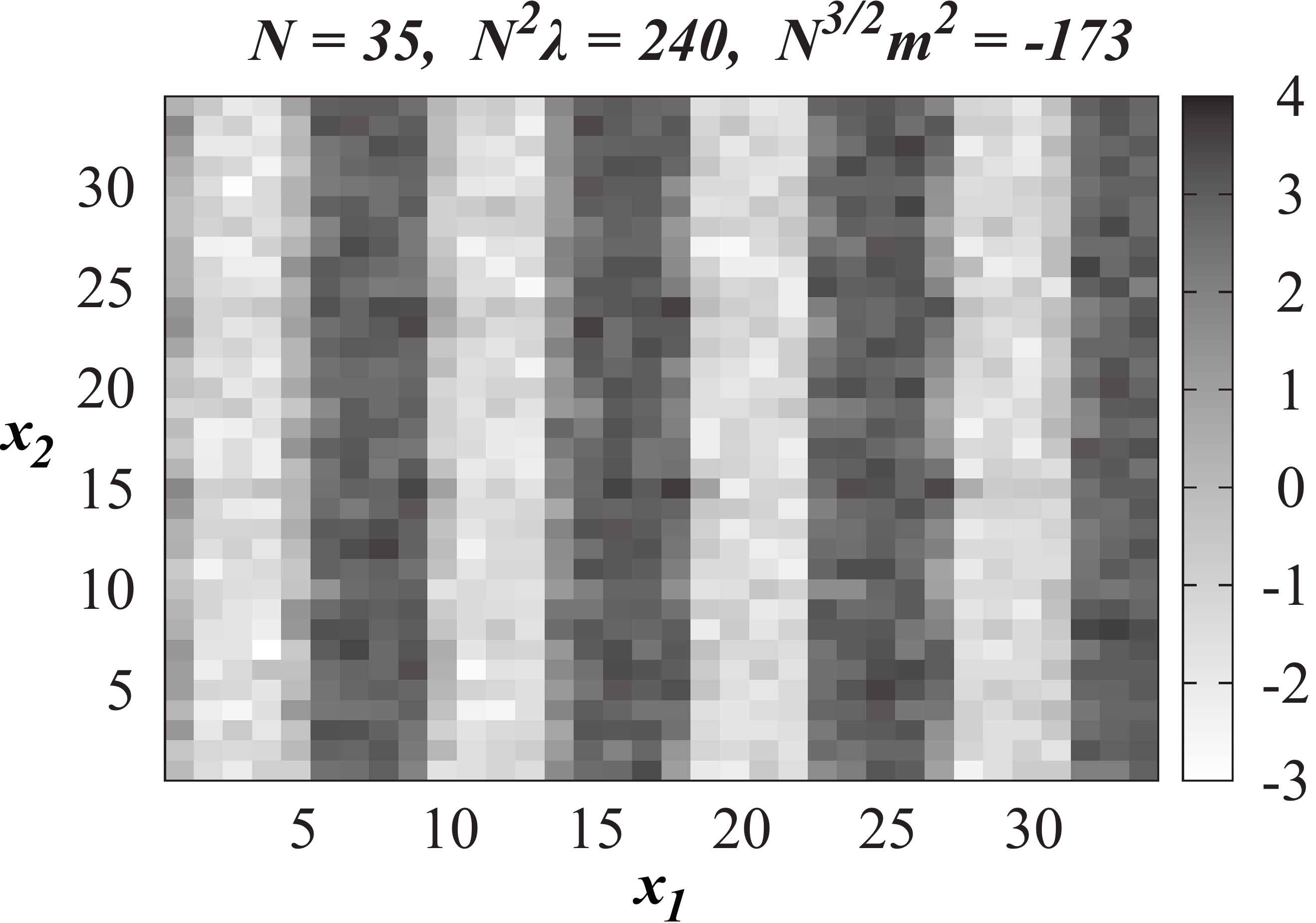}
\caption{\label{fig:striped_phase} A typical configuration in the \emph{striped phase}  obtained in a numerical study of the NC $\phi^4$ theory in two dimensions in ref.~\cite{Mejia-Diaz:2014lza} using the mapping to a matrix model defined in ref.~\cite{Ambjorn:1999ts}.}
\end{center}
\end{figure}

Another recent numerical study of this theory was presented in ref.~\cite{Mejia-Diaz:2014lza}, in which the persistence of the striped phase in the continuum limit was confirmed by accurate extrapolations; as shown in fig.~\ref{fig:striped_phase}, taken from that article, this exotic phase is true to its name, with typical configurations exhibiting a characteristic striped pattern.

Related numerical studies have also been carried out on the fuzzy disc~\cite{Lizzi:2012xy} (based on the construction presented in ref.~\cite{Lizzi:2003ru}), in a tridimensional version of the model~\cite{Bietenholz:2004xs, Medina:2007nv}, at finite temperature~\cite{Das:2007gm},
in a multi-trace formulation~\cite{Ydri:2014uaa, Ydri:2015vba}, et c.\footnote{A different type of study, in which Monte Carlo simulations were used to probe a space of random fuzzy geometries, was recently presented in ref.~\cite{Barrett:2015foa}.} The results of these numerical simulations can be compared with several analytical studies, including refs.~\cite{Minwalla:1999px, Chu:2001xi, Castorina:2003zv, Vaidya:2003ew, Steinacker:2005wj, Castorina:2007za, O'Connor:2007ea, Ihl:2010bu, Nair:2011ux, Kawamoto:2012ng, Polychronakos:2013nca, Tekel:2013vz, Saemann:2014pca, Tekel:2014bta, Rea:2015wta, Ydri:2015zsa, Ydri:2015yta, Tekel:2015zga, Tekel:2015uza}.

\subsection{Results for NC gauge theories}
\label{subsec:results_gauge}

Analytical studies of gauge theories in NC spaces cover a \emph{huge} body of literature: historically, the first example dates back to the first half of the past century~\cite{Snyder:1947nq}. During the past quarter-century, a major research line in these works has been the generalization of the Standard Model of particle physics to a NC framework---although this is only one of the motivations to study NC gauge theories. A very incomplete list of articles addressing problems in this research area includes refs.~\cite{Connes:1990qp, Chamseddine:1992nx, CarowWatamura:1998jn, Martin:1999aq, Madore:2000en, Matusis:2000jf, Iso:2001mg, Jurco:2001my, Chaichian:2001py, Jurco:2001rq, Calmet:2001na, Aschieri:2002mc, Behr:2002wx, Hinchliffe:2002km, Steinacker:2003sd, Grosse:2004wm, Benczik:2005bh, Aschieri:2006uw, Buric:2006wm, Steinacker:2007ay, Chamseddine:2007hz, Grosse:2008xr}.

\begin{figure}
\begin{center}
\includegraphics*[width=0.48\textwidth]{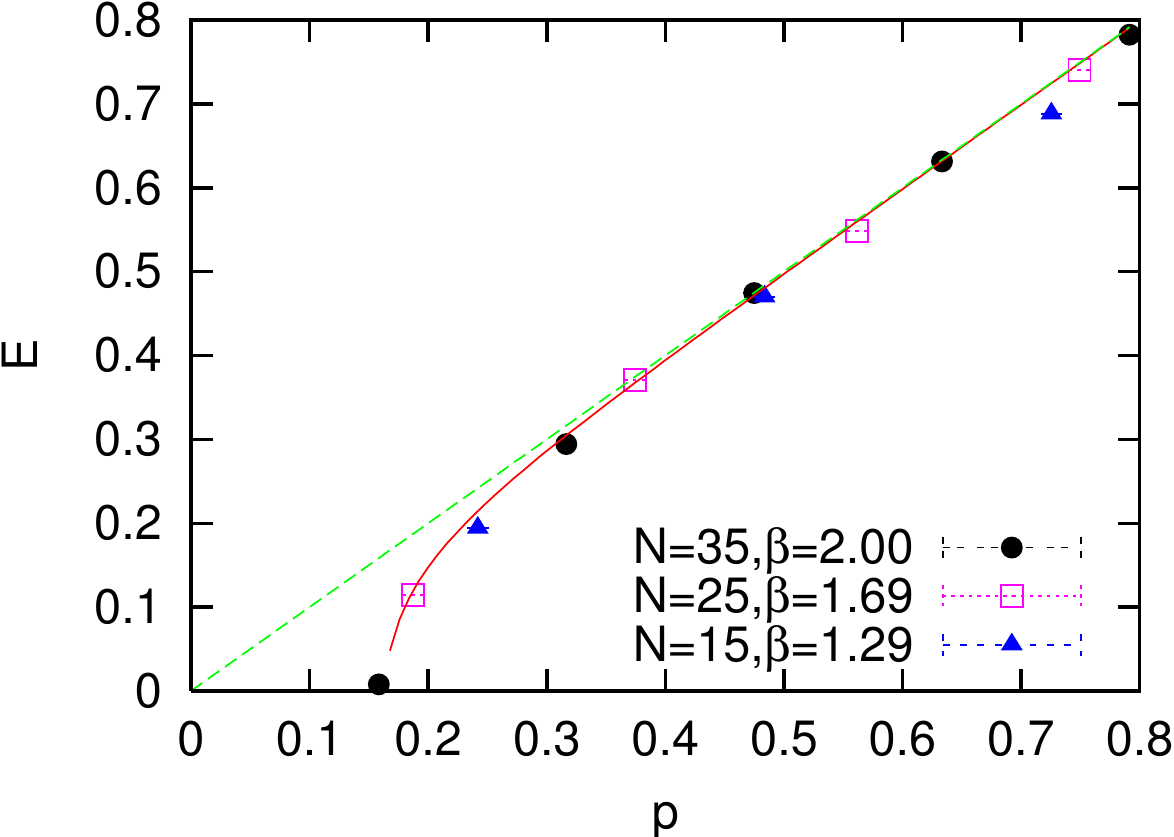}
\caption{\label{fig:photon_dispersion_relation} The photon dispersion relation obtained in numerical simulations of NC QED~\cite{Bietenholz:2006cz} reveals non-linear behavior at small momenta, in agreement with one-loop perturbative predictions~\cite{Hayakawa:1999yt, Martin:2000bk, VanRaamsdonk:2001jd}.}
\end{center}
\end{figure}

Numerical studies of gauge theories in NC spaces, on the other hand, have a relatively recent history~\cite{Bietenholz:2002ch, Azuma:2004zq, CastroVillarreal:2004vh, O'Connor:2006wv, Bietenholz:2006cz, DelgadilloBlando:2007vx, Azeyanagi:2008bk, Spisso:2011wy, O'Connor:2013rla}. An example of results of these studies is shown in fig.~\ref{fig:photon_dispersion_relation}, taken from ref.~\cite{Bietenholz:2006cz}, in which the photon dispersion relation obtained from numerical simulations of NC QED is displayed: the deviations from linear behavior (which are compatible with expectations derived analytically~\cite{Hayakawa:1999yt, Martin:2000bk, VanRaamsdonk:2001jd}) encode an interesting New~Physics signature, which, as discussed in refs.~\cite{Bluemer:2009zf, Szabo:2009tn}, could be potentially observable in ultra-high-energy cosmic rays.

\subsection{Results for supersymmetric models}
\label{subsec:results_supersymmetry}

An interesting recent development in the research on NC theories regularized by means of finite matrices (and in numerical studies thereof) is based on the observation that closely related techniques can be applied for numerical simulations of dimensionally reduced $\mathcal{N}=4$ supersymmetric Yang-Mills theory~\cite{Banks:1996vh}, which describes the dynamics of $N$ D0-branes in type IIA superstring theory~\cite{Witten:1995ex}:
\begin{equation}
S = \frac{N}{2 \lambda} \int_{0}^{\beta}  \dd \tau \, \tr \left\{  \left( D_\tau X_i \right)^2 - \frac{1}{2} \left[ X_i , X_j\right]^2  + \Psi_\alpha D_\tau \Psi _\alpha - \Psi_\alpha \left( \gamma_i \right)_{\alpha\beta} \left[ X_i , \Psi_\beta \right]  \right\}.
\end{equation}
The extension to three or four spacetime dimensions can be achieved, by invoking arguments of large-$N$ volume independence~\cite{Ishiki:2008te}.

\begin{figure}
\begin{center}
\includegraphics*[width=0.48\textwidth]{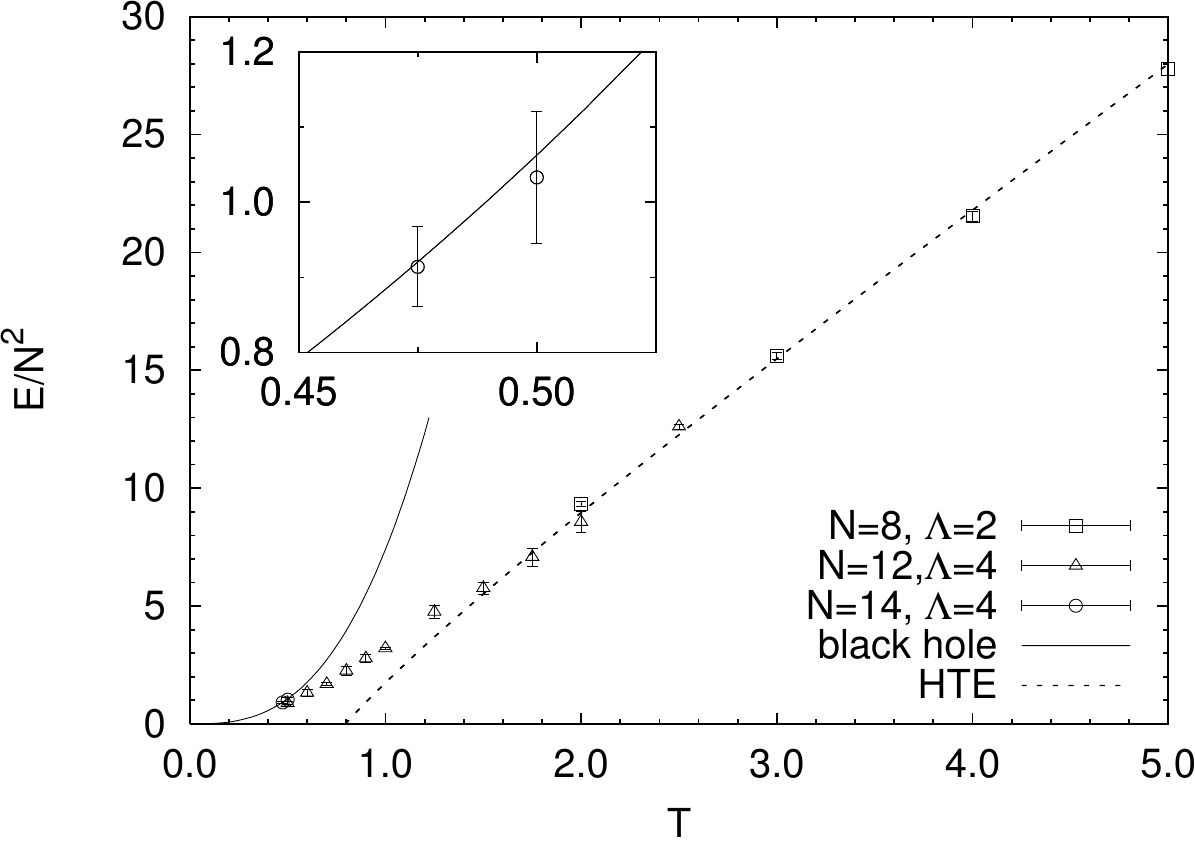} \hfill \includegraphics*[width=0.47\textwidth]{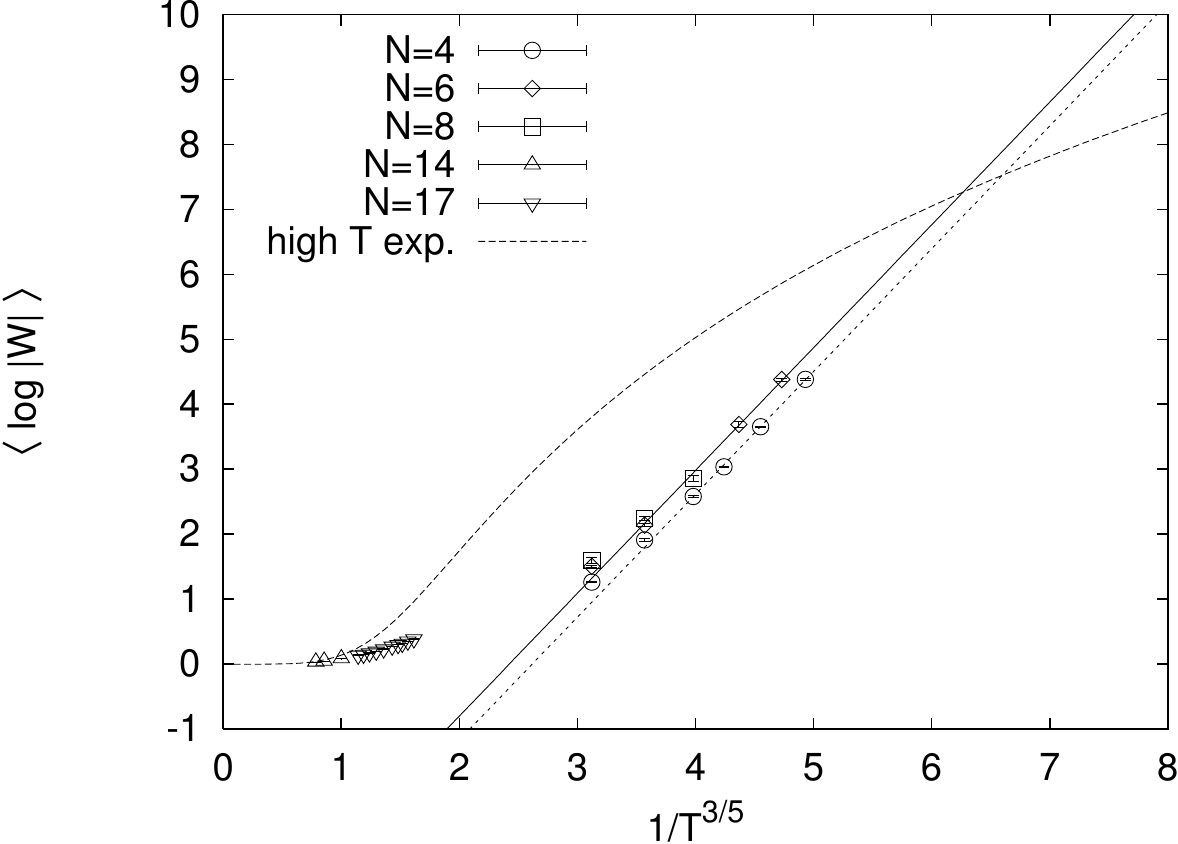}
\caption{\label{fig:susy_QM} Examples of results from numerical simulations of supersymmetric matrix models. The left-hand-side panel shows the energy (normalized by $N^2$) in dimensionally-reduced $\mathcal{N}=4$ supersymmetric Yang-Mills theory, against the temperature $T$~\cite{Anagnostopoulos:2007fw}, in comparison with a high-temperature expansion~\cite{Kawahara:2007ib} and with the classical-supergravity prediction~\cite{Klebanov:1996un}. The right-hand-side panel shows the logarithm of Wilson-Polyakov loop~\cite{Rey:1998ik} in dimensionally-reduced $\mathcal{N}=4$ supersymmetric Yang-Mills theory, against $T^{-3/5}$~\cite{Hanada:2008gy} in comparison with analytical predictions from a high-temperature expansion~\cite{Kawahara:2007ib} and with the expectation in the classical-supergravity limit given by eq.~(\protect\ref{Wilson-Polyakov_loop_sugra}).}
\end{center}
\end{figure}

This approach has been successfully pursued in a number of studies~\cite{Ambjorn:2000bf, Ambjorn:2000dx, Ambjorn:2001xs, Anagnostopoulos:2005cy, Hanada:2007ti, Anagnostopoulos:2007fw, Kawahara:2007eu, Catterall:2008yz, Hanada:2009ne, Hanada:2012si, Anagnostopoulos:2013xga, Hanada:2013rga, Filev:2015hia, Filev:2015cmz}. As an example of results, the left-hand-side panel of fig.~\ref{fig:susy_QM} shows the dependence of the energy on the temperature $T$ in dimensionally reduced, maximally supersymmetric Yang-Mills theory, obtained in the Monte Carlo study reported in ref.~\cite{Anagnostopoulos:2007fw}: at high temperature, the numerical results are in agreement with the expectations from an analytical expansion presented in ref.~\cite{Kawahara:2007ib}, while at low temperatures they can be successfully compared with the classical-supergravity prediction derived in ref.~\cite{Klebanov:1996un}. The right-hand-side panel of fig.~\ref{fig:susy_QM}, instead, shows the numerical results, obtained in ref.~\cite{Hanada:2008gy} using the same formulation of the theory, for the logarithm of Wilson-Polyakov loop introduced in ref.~\cite{Rey:1998ik}: this quantity is plotted against the temperature (to the power $-3/5$) and is compared with the high-temperature expansion derived in ref.~\cite{Kawahara:2007ib} and with the classical-supergravity prediction, which is given by
\begin{equation}
\label{Wilson-Polyakov_loop_sugra}
\langle \ln W \rangle = \sqrt[5]{\frac{120 \pi^2}{49}} \left( \frac{T}{\sqrt[3]{\lambda}} \right)^{-3/5} + \dots .
\end{equation}

\section{Summary and outlook}
\label{sec:summary}

Numerical simulations provide a controlled and systematically improvable tool to study NC QFT from first principles. Not only can they be used to cross-check exact analytical results, but also (more importantly) to determine the validity range of analytical calculations that involve some form of approximation, or to provide guidance to understand problems which are harder to tackle analytically.

As we discussed above, typical Monte~Carlo simulations of NC field theories can be successfully performed with modest computational resources, and the studies of this type that several groups have been carrying out during the past few years have led to many interesting results, for various NC theories.

Interestingly, some techniques devised for simulations of NC QFT can also be applied to field theories defined in ordinary, commutative spaces. For certain problems, the numerical implementation of matrix-model regularization methods inspired by NC field theories proves competitive with respect to more conventional (e.g. lattice) approaches: one striking example is provided by the investigation of dimensionally reduced supersymmetric gauge theories, that we briefly reviewed in subsection~\ref{subsec:results_supersymmetry}.

Finally, it is worthwhile noting that some numerical challenges in Monte~Carlo studies of NC field theories are common to lattice QCD, and any progress toward their solution in one research area could potentially lead to very fruitful developments in the other, too. For example, the authors of refs.~\cite{Anagnostopoulos:2001yb, Anagnostopoulos:2010ux} devised a novel factorization method to cope with the numerical ``sign problem'' affecting their model: as is well-known, this notorious computational problem also hinders simulations of lattice QCD at finite baryon density~\cite{deForcrand:2010ys} and of condensed-matter systems~\cite{Loh:1990sp}.

For all of these reasons, a closer interaction of the two communities would be highly desirable.

\acknowledgments
The author warmly thanks the organizers of the Humboldt Kolleg \emph{``Open Problems in Theoretical Physics: The Issue of Quantum Space-Time''} for the invitation to present this review talk, and the Alexander~von~Humboldt-Stiftung/Foundation for financial support.

\bibliography{paper}

\providecommand{\href}[2]{#2}\begingroup\raggedright\begin{thebibliography}{100}

\bibitem{Doplicher:1994tu}
S.~Doplicher, K.~Fredenhagen and J.~E. Roberts, {\it {The Quantum structure of
  space-time at the Planck scale and quantum fields}},  {\em Commun. Math.
  Phys.} {\bf 172} (1995) 187--220,
  [\href{http://xxx.lanl.gov/abs/hep-th/0303037}{{\tt hep-th/0303037}}].

\bibitem{Seiberg:1999vs}
N.~Seiberg and E.~Witten, {\it {String theory and noncommutative geometry}},
  {\em JHEP} {\bf 9909} (1999) 032,
  [\href{http://xxx.lanl.gov/abs/hep-th/9908142}{{\tt hep-th/9908142}}].

\bibitem{Douglas:2001ba}
M.~R. Douglas and N.~A. Nekrasov, {\it {Noncommutative field theory}},  {\em
  Rev.Mod.Phys.} {\bf 73} (2001) 977--1029,
  [\href{http://xxx.lanl.gov/abs/hep-th/0106048}{{\tt hep-th/0106048}}].

\bibitem{Szabo:2001kg}
R.~J. Szabo, {\it {Quantum field theory on noncommutative spaces}},  {\em
  Phys.Rept.} {\bf 378} (2003) 207--299,
  [\href{http://xxx.lanl.gov/abs/hep-th/0109162}{{\tt hep-th/0109162}}].

\bibitem{Minwalla:1999px}
S.~Minwalla, M.~Van~Raamsdonk and N.~Seiberg, {\it {Noncommutative
  perturbative dynamics}},  {\em JHEP} {\bf 02} (2000) 020,
  [\href{http://xxx.lanl.gov/abs/hep-th/9912072}{{\tt hep-th/9912072}}].

\bibitem{Feynman:1948ur}
R.~P. Feynman, {\it {Space-time approach to nonrelativistic quantum
  mechanics}},  {\em Rev. Mod. Phys.} {\bf 20} (1948) 367--387.

\bibitem{Maldacena:1997re}
J.~M. Maldacena, {\it {The Large N limit of superconformal field theories and
  supergravity}},  {\em Adv.Theor.Math.Phys.} {\bf 2} (1998) 231--252,
  [\href{http://xxx.lanl.gov/abs/hep-th/9711200}{{\tt hep-th/9711200}}].

\bibitem{Gubser:1998bc}
S.~Gubser, I.~R. Klebanov and A.~M. Polyakov, {\it {Gauge theory correlators
  from noncritical string theory}},  {\em Phys.Lett.} {\bf B428} (1998)
  105--114, [\href{http://xxx.lanl.gov/abs/hep-th/9802109}{{\tt
  hep-th/9802109}}].

\bibitem{Witten:1998qj}
E.~Witten, {\it {Anti-de Sitter space and holography}},  {\em
  Adv.Theor.Math.Phys.} {\bf 2} (1998) 253--291,
  [\href{http://xxx.lanl.gov/abs/hep-th/9802150}{{\tt hep-th/9802150}}].

\bibitem{Matthews:1954zg}
P.~T. Matthews and A.~Salam, {\it {The Green's functions of quantized fields}},
   {\em Nuovo Cim.} {\bf 12} (1954) 563--565.

\bibitem{Matthews:1955zi}
P.~T. Matthews and A.~Salam, {\it {Propagators of quantized field}},  {\em
  Nuovo Cim.} {\bf 2} (1955) 120--134.

\bibitem{Madore:1991bw}
J.~Madore, {\it {The Fuzzy sphere}},  {\em Class. Quant. Grav.} {\bf 9} (1992)
  69--88.

\bibitem{Martin:2004un}
X.~Martin, {\it {A Matrix phase for the {$\phi^4$} scalar field on the fuzzy
  sphere}},  {\em JHEP} {\bf 04} (2004) 077,
  [\href{http://xxx.lanl.gov/abs/hep-th/0402230}{{\tt hep-th/0402230}}].

\bibitem{Panero:2006bx}
M.~Panero, {\it {Numerical simulations of a non-commutative theory: The Scalar
  model on the fuzzy sphere}},  {\em JHEP} {\bf 0705} (2007) 082,
  [\href{http://xxx.lanl.gov/abs/hep-th/0608202}{{\tt hep-th/0608202}}].

\bibitem{Panero:2006cs}
M.~Panero, {\it {Quantum field theory in a non-commutative space: Theoretical
  predictions and numerical results on the fuzzy sphere}},  {\em SIGMA} {\bf 2}
  (2006) 081, [\href{http://xxx.lanl.gov/abs/hep-th/0609205}{{\tt
  hep-th/0609205}}].

\bibitem{GarciaFlores:2009hf}
F.~Garc{\'{\i}}a~Flores, X.~Martin and D.~O'Connor, {\it {Simulation of a
  scalar field on a fuzzy sphere}},  {\em Int. J. Mod. Phys.} {\bf A24} (2009)
  3917--3944, [\href{http://xxx.lanl.gov/abs/0903.1986}{{\tt
  arXiv:0903.1986}}].

\bibitem{Digal:2011hf}
S.~Digal, T.~R. Govindarajan, K.~S. Gupta and X.~Martin, {\it {Phase structure
  of fuzzy black holes}},  {\em JHEP} {\bf 01} (2012) 027,
  [\href{http://xxx.lanl.gov/abs/1109.4014}{{\tt arXiv:1109.4014}}].

\bibitem{Okuno:2015kuc}
S.~Okuno, M.~Suzuki and A.~Tsuchiya, {\it {Entanglement entropy in scalar
  field theory on the fuzzy sphere}},
  \href{http://xxx.lanl.gov/abs/1512.06484}{{\tt arXiv:1512.06484}}.

\bibitem{Ambjorn:1999ts}
J.~Ambj{\o}rn, Y.~Makeenko, J.~Nishimura and R.~Szabo, {\it {Finite N matrix
  models of noncommutative gauge theory}},  {\em JHEP} {\bf 9911} (1999) 029,
  [\href{http://xxx.lanl.gov/abs/hep-th/9911041}{{\tt hep-th/9911041}}].

\bibitem{GonzalezArroyo:1982ub}
A.~Gonz{\'a}lez-Arroyo and M.~Okawa, {\it {A twisted model for large N lattice
  gauge theory}},  {\em Phys.Lett.} {\bf B120} (1983) 174.

\bibitem{GonzalezArroyo:2010ss}
A.~Gonz{\'a}lez-Arroyo and M.~Okawa, {\it {Large $N$ reduction with the Twisted
  Eguchi-Kawai model}},  {\em JHEP} {\bf 1007} (2010) 043,
  [\href{http://xxx.lanl.gov/abs/1005.1981}{{\tt arXiv:1005.1981}}].

\bibitem{Lucini:2012gg}
B.~Lucini and M.~Panero, {\it {SU(N) gauge theories at large N}},  {\em
  Phys.Rept.} {\bf 526} (2013) 93--163,
  [\href{http://xxx.lanl.gov/abs/1210.4997}{{\tt arXiv:1210.4997}}].

\bibitem{Ambjorn:2002nj}
J.~Ambj{\o}rn and S.~Catterall, {\it {Stripes from (noncommutative) stars}},
  {\em Phys. Lett.} {\bf B549} (2002) 253--259,
  [\href{http://xxx.lanl.gov/abs/hep-lat/0209106}{{\tt hep-lat/0209106}}].

\bibitem{Bietenholz:2002ev}
W.~Bietenholz, F.~Hofheinz and J.~Nishimura, {\it {Noncommutative field
  theories beyond perturbation theory}},  {\em Fortsch. Phys.} {\bf 51} (2003)
  745--752, [\href{http://xxx.lanl.gov/abs/hep-th/0212258}{{\tt
  hep-th/0212258}}].

\bibitem{Mejia-Diaz:2014lza}
H.~Mej{\'{\i}}a-D{\'{\i}}az, W.~Bietenholz and M.~Panero, {\it {The continuum
  phase diagram of the 2d non-commutative $\lambda \phi^4$ model}},  {\em JHEP}
  {\bf 10} (2014) 56, [\href{http://xxx.lanl.gov/abs/1403.3318}{{\tt
  arXiv:1403.3318}}].

\bibitem{Wilson:1974sk}
K.~G. Wilson, {\it {Confinement of Quarks}},  {\em Phys.Rev.} {\bf D10} (1974)
  2445--2459.

\bibitem{Grosse:1995pr}
H.~Grosse, C.~Klim\v{c}\'{\i}k and P.~Pre{\v{s}}najder, {\it {Field theory on
  a supersymmetric lattice}},  {\em Commun. Math. Phys.} {\bf 185} (1997)
  155--175, [\href{http://xxx.lanl.gov/abs/hep-th/9507074}{{\tt
  hep-th/9507074}}].

\bibitem{Balachandran:2005ew}
A.~Balachandran, S.~K{\"u}rk{\c{c}}{\"{u}}o{\v{g}}lu and S.~Vaidya, {\em
  {Lectures on fuzzy and fuzzy SUSY physics}}.
\newblock World Scientific, Singapore, 2007.

\bibitem{Grosse:1994ed}
H.~Grosse and P.~Pre{\v{s}}najder, {\it {The Dirac operator on the fuzzy
  sphere}},  {\em Lett. Math. Phys.} {\bf 33} (1995) 171--182.

\bibitem{Grosse:1995jt}
H.~Grosse, C.~Klim\v{c}\'{\i}k and P.~Pre{\v{s}}najder, {\it {Topologically
  nontrivial field configurations in noncommutative geometry}},  {\em Commun.
  Math. Phys.} {\bf 178} (1996) 507--526,
  [\href{http://xxx.lanl.gov/abs/hep-th/9510083}{{\tt hep-th/9510083}}].

\bibitem{Balachandran:1999qu}
A.~P. Balachandran, T.~R. Govindarajan and B.~Ydri, {\it {The Fermion doubling
  problem and noncommutative geometry}},  {\em Mod. Phys. Lett.} {\bf A15}
  (2000) 1279, [\href{http://xxx.lanl.gov/abs/hep-th/9911087}{{\tt
  hep-th/9911087}}].

\bibitem{Balachandran:2003ay}
A.~P. Balachandran and G.~Immirzi, {\it {The Fuzzy Ginsparg-Wilson algebra: A
  Solution of the fermion doubling problem}},  {\em Phys. Rev.} {\bf D68}
  (2003) 065023, [\href{http://xxx.lanl.gov/abs/hep-th/0301242}{{\tt
  hep-th/0301242}}].

\bibitem{Dolan:2007uf}
B.~P. Dolan, I.~Huet, S.~Murray and D.~O'Connor, {\it {A Universal Dirac
  operator and noncommutative spin bundles over fuzzy complex projective
  spaces}},  {\em JHEP} {\bf 03} (2008) 029,
  [\href{http://xxx.lanl.gov/abs/0711.1347}{{\tt arXiv:0711.1347}}].

\bibitem{Nielsen:1980rz}
H.~B. Nielsen and M.~Ninomiya, {\it {Absence of Neutrinos on a Lattice. 1.
  Proof by Homotopy Theory}},  {\em Nucl. Phys.} {\bf B185} (1981) 20.
  [Erratum: Nucl. Phys.B195,541(1982)].

\bibitem{Nielsen:1981xu}
H.~B. Nielsen and M.~Ninomiya, {\it {Absence of Neutrinos on a Lattice. 2.
  Intuitive Topological Proof}},  {\em Nucl. Phys.} {\bf B193} (1981) 173.

\bibitem{Wilson:1975hf}
K.~G. Wilson, {\it {Quarks: From Paradox to Myth}},  {\em Subnucl. Ser.} {\bf
  13} (1977) 13--32.

\bibitem{Kaplan:1992bt}
D.~B. Kaplan, {\it {A Method for simulating chiral fermions on the lattice}},
  {\em Phys.Lett.} {\bf B288} (1992) 342--347,
  [\href{http://xxx.lanl.gov/abs/hep-lat/9206013}{{\tt hep-lat/9206013}}].

\bibitem{Narayanan:1994gw}
R.~Narayanan and H.~Neuberger, {\it {A Construction of lattice chiral gauge
  theories}},  {\em Nucl.Phys.} {\bf B443} (1995) 305--385,
  [\href{http://xxx.lanl.gov/abs/hep-th/9411108}{{\tt hep-th/9411108}}].

\bibitem{Neuberger:1997fp}
H.~Neuberger, {\it {Exactly massless quarks on the lattice}},  {\em Phys.Lett.}
  {\bf B417} (1998) 141--144,
  [\href{http://xxx.lanl.gov/abs/hep-lat/9707022}{{\tt hep-lat/9707022}}].

\bibitem{Nishimura:2009xm}
J.~Nishimura, {\it {Non-lattice simulation of supersymmetric gauge theories as
  a probe to quantum black holes and strings}},  {\em PoS} {\bf Lattice 2009}
  (2009) 016, [\href{http://xxx.lanl.gov/abs/0912.0327}{{\tt
  arXiv:0912.0327}}].

\bibitem{Joseph:2015xwa}
A.~Joseph, {\it {Review of Lattice Supersymmetry and Gauge-Gravity Duality}},
  {\em Int. J. Mod. Phys.} {\bf A30} (2015), no.~27 1530054,
  [\href{http://xxx.lanl.gov/abs/1509.01440}{{\tt arXiv:1509.01440}}].

\bibitem{Bergner:2013nwa}
G.~Bergner, I.~Montvay, G.~M{\"u}nster, U.~D. {\"O}zugurel and D.~Sandbrink,
  {\it {Towards the spectrum of low-lying particles in supersymmetric
  Yang-Mills theory}},  {\em JHEP} {\bf 11} (2013) 061,
  [\href{http://xxx.lanl.gov/abs/1304.2168}{{\tt arXiv:1304.2168}}].

\bibitem{Catterall:2009it}
S.~Catterall, D.~B. Kaplan and M.~{\"U}nsal, {\it {Exact lattice
  supersymmetry}},  {\em Phys.Rept.} {\bf 484} (2009) 71--130,
  [\href{http://xxx.lanl.gov/abs/0903.4881}{{\tt arXiv:0903.4881}}].

\bibitem{Kronfeld:2012uk}
A.~S. Kronfeld, {\it {Twenty-first Century Lattice Gauge Theory: Results from
  the QCD Lagrangian}},  {\em Ann.Rev.Nucl.Part.Sci.} {\bf 62} (2012) 265--284,
  [\href{http://xxx.lanl.gov/abs/1203.1204}{{\tt arXiv:1203.1204}}].

\bibitem{Panero:2012qx}
M.~Panero, {\it {Recent results in large-$N$ lattice gauge theories}},  {\em
  PoS} {\bf Lattice 2012} (2012) 010,
  [\href{http://xxx.lanl.gov/abs/1210.5510}{{\tt arXiv:1210.5510}}].

\bibitem{Lucini:2013qja}
B.~Lucini and M.~Panero, {\it {Introductory lectures to
  large-{$\scriptsize{N}$} QCD phenomenology and lattice results}},  {\em
  Prog.Part.Nucl.Phys.} {\bf 75} (2014) 1--40,
  [\href{http://xxx.lanl.gov/abs/1309.3638}{{\tt arXiv:1309.3638}}].

\bibitem{DelDebbio:2010zz}
L.~Del~Debbio, {\it {The conformal window on the lattice}},  {\em PoS} {\bf
  Lattice 2010} (2010) 004.

\bibitem{Giedt:2012it}
J.~Giedt, {\it {Lattice gauge theory and physics beyond the standard model}},
  {\em PoS} {\bf Lattice 2012} (2012) 006.

\bibitem{Bietenholz:2014sza}
W.~Bietenholz, F.~Hofheinz, H.~Mej{\'{\i}}a-D{\'{\i}}az and M.~Panero, {\it
  {Scalar fields in a non-commutative space}},  in {\em {14th Mexican Workshop
  on Particles and Fields (MWPF 2013) Oaxaca, Mexico, November 25-29, 2013}},
  vol.~651, p.~012003, 2015.
\newblock \href{http://xxx.lanl.gov/abs/1402.4420}{{\tt arXiv:1402.4420}}.

\bibitem{Gubser:2000cd}
S.~S. Gubser and S.~L. Sondhi, {\it {Phase structure of noncommutative scalar
  field theories}},  {\em Nucl. Phys.} {\bf B605} (2001) 395--424,
  [\href{http://xxx.lanl.gov/abs/hep-th/0006119}{{\tt hep-th/0006119}}].

\bibitem{Lizzi:2012xy}
F.~Lizzi and B.~Spisso, {\it {Noncommutative Field Theory: Numerical Analysis
  with the Fuzzy Disc}},  {\em Int. J. Mod. Phys.} {\bf A27} (2012) 1250137,
  [\href{http://xxx.lanl.gov/abs/1207.4998}{{\tt arXiv:1207.4998}}].

\bibitem{Lizzi:2003ru}
F.~Lizzi, P.~Vitale and A.~Zampini, {\it {The Fuzzy disc}},  {\em JHEP} {\bf
  08} (2003) 057, [\href{http://xxx.lanl.gov/abs/hep-th/0306247}{{\tt
  hep-th/0306247}}].

\bibitem{Bietenholz:2004xs}
W.~Bietenholz, F.~Hofheinz and J.~Nishimura, {\it {Phase diagram and
  dispersion relation of the noncommutative lambda {$\phi^4$} model in {$d =
  3$}}},  {\em JHEP} {\bf 06} (2004) 042,
  [\href{http://xxx.lanl.gov/abs/hep-th/0404020}{{\tt hep-th/0404020}}].

\bibitem{Medina:2007nv}
J.~Medina, W.~Bietenholz and D.~O'Connor, {\it {Probing the fuzzy sphere
  regularisation in simulations of the 3d lambda {$\phi^4$} model}},  {\em
  JHEP} {\bf 04} (2008) 041, [\href{http://xxx.lanl.gov/abs/0712.3366}{{\tt
  arXiv:0712.3366}}].

\bibitem{Das:2007gm}
C.~R. Das, S.~Digal and T.~R. Govindarajan, {\it {Finite temperature phase
  transition of a single scalar field on a fuzzy sphere}},  {\em Mod. Phys.
  Lett.} {\bf A23} (2008) 1781--1791,
  [\href{http://xxx.lanl.gov/abs/0706.0695}{{\tt arXiv:0706.0695}}].

\bibitem{Ydri:2014uaa}
B.~Ydri, {\it {A Multitrace Approach to Noncommutative {$\Phi_2^4$}}},
  \href{http://xxx.lanl.gov/abs/1410.4881}{{\tt arXiv:1410.4881}}.

\bibitem{Ydri:2015vba}
B.~Ydri, K.~Ramda and A.~Rouag, {\it {Phase diagrams of the multitrace quartic
  matrix models of noncommutative {$\Phi^4$}}},
  \href{http://xxx.lanl.gov/abs/1509.03726}{{\tt arXiv:1509.03726}}.

\bibitem{Barrett:2015foa}
J.~W. Barrett and L.~Glaser, {\it {Monte Carlo simulations of random
  non-commutative geometries}},  \href{http://xxx.lanl.gov/abs/1510.01377}{{\tt
  arXiv:1510.01377}}.

\bibitem{Chu:2001xi}
C.-S. Chu, J.~Madore and H.~Steinacker, {\it {Scaling limits of the fuzzy
  sphere at one loop}},  {\em JHEP} {\bf 08} (2001) 038,
  [\href{http://xxx.lanl.gov/abs/hep-th/0106205}{{\tt hep-th/0106205}}].

\bibitem{Castorina:2003zv}
P.~Castorina and D.~Zappal{\`a}, {\it {Nonuniform symmetry breaking in
  noncommutative {$\lambda \phi^4$} theory}},  {\em Phys. Rev.} {\bf D68}
  (2003) 065008, [\href{http://xxx.lanl.gov/abs/hep-th/0303030}{{\tt
  hep-th/0303030}}].

\bibitem{Vaidya:2003ew}
S.~Vaidya and B.~Ydri, {\it {On the origin of the UV-IR mixing in
  noncommutative matrix geometry}},  {\em Nucl. Phys.} {\bf B671} (2003)
  401--431, [\href{http://xxx.lanl.gov/abs/hep-th/0305201}{{\tt
  hep-th/0305201}}].

\bibitem{Steinacker:2005wj}
H.~Steinacker, {\it {A Non-perturbative approach to non-commutative scalar
  field theory}},  {\em JHEP} {\bf 03} (2005) 075,
  [\href{http://xxx.lanl.gov/abs/hep-th/0501174}{{\tt hep-th/0501174}}].

\bibitem{Castorina:2007za}
P.~Castorina and D.~Zappal{\`a}, {\it {Spontaneous breaking of translational
  invariance in non-commutative {$\lambda \phi^4$} theory in two dimensions}},
  {\em Phys. Rev.} {\bf D77} (2008) 027703,
  [\href{http://xxx.lanl.gov/abs/0711.2659}{{\tt arXiv:0711.2659}}].

\bibitem{O'Connor:2007ea}
D.~O'Connor and C.~S{\"a}mann, {\it {Fuzzy Scalar Field Theory as a Multitrace
  Matrix Model}},  {\em JHEP} {\bf 08} (2007) 066,
  [\href{http://xxx.lanl.gov/abs/0706.2493}{{\tt arXiv:0706.2493}}].

\bibitem{Ihl:2010bu}
M.~Ihl, C.~Sachse and C.~S{\"a}mann, {\it {Fuzzy Scalar Field Theory as Matrix
  Quantum Mechanics}},  {\em JHEP} {\bf 03} (2011) 091,
  [\href{http://xxx.lanl.gov/abs/1012.3568}{{\tt arXiv:1012.3568}}].

\bibitem{Nair:2011ux}
V.~P. Nair, A.~P. Polychronakos and J.~Tekel, {\it {Fuzzy spaces and new
  random matrix ensembles}},  {\em Phys. Rev.} {\bf D85} (2012) 045021,
  [\href{http://xxx.lanl.gov/abs/1109.3349}{{\tt arXiv:1109.3349}}].

\bibitem{Kawamoto:2012ng}
S.~Kawamoto, T.~Kuroki and D.~Tomino, {\it {Renormalization group approach to
  matrix models via noncommutative space}},  {\em JHEP} {\bf 08} (2012) 168,
  [\href{http://xxx.lanl.gov/abs/1206.0574}{{\tt arXiv:1206.0574}}].

\bibitem{Polychronakos:2013nca}
A.~P. Polychronakos, {\it {Effective action and phase transitions of scalar
  field on the fuzzy sphere}},  {\em Phys. Rev.} {\bf D88} (2013) 065010,
  [\href{http://xxx.lanl.gov/abs/1306.6645}{{\tt arXiv:1306.6645}}].

\bibitem{Tekel:2013vz}
J.~Tekel, {\it {Random matrix approach to scalar fields on fuzzy spaces}},
  {\em Phys. Rev.} {\bf D87} (2013), no.~8 085015,
  [\href{http://xxx.lanl.gov/abs/1301.2154}{{\tt arXiv:1301.2154}}].

\bibitem{Saemann:2014pca}
C.~S{\"a}mann, {\it {Bootstrapping Fuzzy Scalar Field Theory}},  {\em JHEP}
  {\bf 04} (2015) 044, [\href{http://xxx.lanl.gov/abs/1412.6255}{{\tt
  arXiv:1412.6255}}].

\bibitem{Tekel:2014bta}
J.~Tekel, {\it {Uniform order phase and phase diagram of scalar field theory on
  fuzzy $\mathbb C P^n$}},  {\em JHEP} {\bf 10} (2014) 144,
  [\href{http://xxx.lanl.gov/abs/1407.4061}{{\tt arXiv:1407.4061}}].

\bibitem{Rea:2015wta}
S.~Rea and C.~S{\"a}mann, {\it {The Phase Diagram of Scalar Field Theory on the
  Fuzzy Disc}},  {\em JHEP} {\bf 11} (2015) 115,
  [\href{http://xxx.lanl.gov/abs/1507.05978}{{\tt arXiv:1507.05978}}].

\bibitem{Ydri:2015zsa}
B.~Ydri, A.~Rouag and K.~Ramda, {\it {Emergent geometry from random multitrace
  matrix models}},  \href{http://xxx.lanl.gov/abs/1509.03572}{{\tt
  arXiv:1509.03572}}.

\bibitem{Ydri:2015yta}
B.~Ydri, R.~Ahmim and A.~Bouchareb, {\it {Wilson RG of Noncommutative
  $\Phi_{4}^4$}},  {\em Int. J. Mod. Phys.} {\bf A30} (2015), no.~33 1550195,
  [\href{http://xxx.lanl.gov/abs/1509.03605}{{\tt arXiv:1509.03605}}].

\bibitem{Tekel:2015zga}
J.~Tekel, {\it {Matrix model approximations of fuzzy scalar field theories and
  their phase diagrams}},  {\em JHEP} {\bf 12} (2015) 176,
  [\href{http://xxx.lanl.gov/abs/1510.07496}{{\tt arXiv:1510.07496}}].

\bibitem{Tekel:2015uza}
J.~Tekel, {\it {Phase structure of fuzzy field theories and multitrace matrix
  models}},  {\em Acta Phys. Slov.} {\bf 65} (2015) 369,
  [\href{http://xxx.lanl.gov/abs/1512.00689}{{\tt arXiv:1512.00689}}].

\bibitem{Snyder:1947nq}
H.~S. Snyder, {\it {The Electromagnetic Field in Quantized Space-Time}},  {\em
  Phys. Rev.} {\bf 72} (1947) 68--71.

\bibitem{Connes:1990qp}
A.~Connes and J.~Lott, {\it {Particle Models and Noncommutative Geometry
  (Expanded Version)}},  {\em Nucl. Phys. Proc. Suppl.} {\bf 18B} (1991)
  29--47.

\bibitem{Chamseddine:1992nx}
A.~H. Chamseddine, G.~Felder and J.~Fr{\"o}hlich, {\it {Grand unification in
  noncommutative geometry}},  {\em Nucl. Phys.} {\bf B395} (1993) 672--700,
  [\href{http://xxx.lanl.gov/abs/hep-ph/9209224}{{\tt hep-ph/9209224}}].

\bibitem{CarowWatamura:1998jn}
U.~Carow-Watamura and S.~Watamura, {\it {Noncommutative geometry and gauge
  theory on fuzzy sphere}},  {\em Commun. Math. Phys.} {\bf 212} (2000)
  395--413, [\href{http://xxx.lanl.gov/abs/hep-th/9801195}{{\tt
  hep-th/9801195}}].

\bibitem{Martin:1999aq}
C.~P. Mart{\'{\i}}n and D.~S{\'a}nchez-Ruiz, {\it {The one loop UV divergent
  structure of U(1) Yang-Mills theory on noncommutative {$R^4$}}},  {\em Phys.
  Rev. Lett.} {\bf 83} (1999) 476--479,
  [\href{http://xxx.lanl.gov/abs/hep-th/9903077}{{\tt hep-th/9903077}}].

\bibitem{Madore:2000en}
J.~Madore, S.~Schraml, P.~Schupp and J.~Wess, {\it {Gauge theory on
  noncommutative spaces}},  {\em Eur. Phys. J.} {\bf C16} (2000) 161--167,
  [\href{http://xxx.lanl.gov/abs/hep-th/0001203}{{\tt hep-th/0001203}}].

\bibitem{Matusis:2000jf}
A.~Matusis, L.~Susskind and N.~Toumbas, {\it {The IR / UV connection in the
  noncommutative gauge theories}},  {\em JHEP} {\bf 12} (2000) 002,
  [\href{http://xxx.lanl.gov/abs/hep-th/0002075}{{\tt hep-th/0002075}}].

\bibitem{Iso:2001mg}
S.~Iso, Y.~Kimura, K.~Tanaka and K.~Wakatsuki, {\it {Noncommutative gauge
  theory on fuzzy sphere from matrix model}},  {\em Nucl. Phys.} {\bf B604}
  (2001) 121--147, [\href{http://xxx.lanl.gov/abs/hep-th/0101102}{{\tt
  hep-th/0101102}}].

\bibitem{Jurco:2001my}
B.~Jur{\v{c}}o, P.~Schupp and J.~Wess, {\it {NonAbelian noncommutative gauge
  theory via noncommutative extra dimensions}},  {\em Nucl. Phys.} {\bf B604}
  (2001) 148--180, [\href{http://xxx.lanl.gov/abs/hep-th/0102129}{{\tt
  hep-th/0102129}}].

\bibitem{Chaichian:2001py}
M.~Chaichian, P.~Pre{\v{s}}najder, M.~M. Sheikh-Jabbari and A.~Tureanu, {\it
  {Noncommutative standard model: Model building}},  {\em Eur. Phys. J.} {\bf
  C29} (2003) 413--432, [\href{http://xxx.lanl.gov/abs/hep-th/0107055}{{\tt
  hep-th/0107055}}].

\bibitem{Jurco:2001rq}
B.~Jur{\v{c}}o, L.~M{\"o}ller, S.~Schraml, P.~Schupp and J.~Wess, {\it
  {Construction of nonAbelian gauge theories on noncommutative spaces}},  {\em
  Eur. Phys. J.} {\bf C21} (2001) 383--388,
  [\href{http://xxx.lanl.gov/abs/hep-th/0104153}{{\tt hep-th/0104153}}].

\bibitem{Calmet:2001na}
X.~Calmet, B.~Jur{\v{c}}o, P.~Schupp, J.~Wess and M.~Wohlgenannt, {\it {The
  Standard model on noncommutative space-time}},  {\em Eur. Phys. J.} {\bf C23}
  (2002) 363--376, [\href{http://xxx.lanl.gov/abs/hep-ph/0111115}{{\tt
  hep-ph/0111115}}].

\bibitem{Aschieri:2002mc}
P.~Aschieri, B.~Jur{\v{c}}o, P.~Schupp and J.~Wess, {\it {Noncommutative GUTs,
  standard model and C,P,T}},  {\em Nucl. Phys.} {\bf B651} (2003) 45--70,
  [\href{http://xxx.lanl.gov/abs/hep-th/0205214}{{\tt hep-th/0205214}}].

\bibitem{Behr:2002wx}
W.~Behr, N.~G. Deshpande, G.~Duplan{\v{c}}i{\'c}, P.~Schupp, J.~Trampeti{\'c}
  and J.~Wess, {\it {The $ Z \to \gamma \gamma, g g $ decays in the
  noncommutative standard model}},  {\em Eur. Phys. J.} {\bf C29} (2003)
  441--446, [\href{http://xxx.lanl.gov/abs/hep-ph/0202121}{{\tt
  hep-ph/0202121}}].

\bibitem{Hinchliffe:2002km}
I.~Hinchliffe, N.~Kersting and Y.~L. Ma, {\it {Review of the phenomenology of
  noncommutative geometry}},  {\em Int. J. Mod. Phys.} {\bf A19} (2004)
  179--204, [\href{http://xxx.lanl.gov/abs/hep-ph/0205040}{{\tt
  hep-ph/0205040}}].

\bibitem{Steinacker:2003sd}
H.~Steinacker, {\it {Quantized gauge theory on the fuzzy sphere as random
  matrix model}},  {\em Nucl. Phys.} {\bf B679} (2004) 66--98,
  [\href{http://xxx.lanl.gov/abs/hep-th/0307075}{{\tt hep-th/0307075}}].

\bibitem{Grosse:2004wm}
H.~Grosse and H.~Steinacker, {\it {Finite gauge theory on fuzzy {CP$^2$}}},
  {\em Nucl. Phys.} {\bf B707} (2005) 145--198,
  [\href{http://xxx.lanl.gov/abs/hep-th/0407089}{{\tt hep-th/0407089}}].

\bibitem{Benczik:2005bh}
S.~Benczik, L.~N. Chang, D.~Minic and T.~Takeuchi, {\it {The Hydrogen atom
  with minimal length}},  {\em Phys. Rev.} {\bf A72} (2005) 012104,
  [\href{http://xxx.lanl.gov/abs/hep-th/0502222}{{\tt hep-th/0502222}}].

\bibitem{Aschieri:2006uw}
P.~Aschieri, T.~Grammatikopoulos, H.~Steinacker and G.~Zoupanos, {\it
  {Dynamical generation of fuzzy extra dimensions, dimensional reduction and
  symmetry breaking}},  {\em JHEP} {\bf 09} (2006) 026,
  [\href{http://xxx.lanl.gov/abs/hep-th/0606021}{{\tt hep-th/0606021}}].

\bibitem{Buric:2006wm}
M.~Buri{\'c}, V.~Radovanovi{\'c} and J.~Trampeti{\'c}, {\it {The One-loop
  renormalization of the gauge sector in the noncommutative standard model}},
  {\em JHEP} {\bf 03} (2007) 030,
  [\href{http://xxx.lanl.gov/abs/hep-th/0609073}{{\tt hep-th/0609073}}].

\bibitem{Steinacker:2007ay}
H.~Steinacker and G.~Zoupanos, {\it {Fermions on spontaneously generated
  spherical extra dimensions}},  {\em JHEP} {\bf 09} (2007) 017,
  [\href{http://xxx.lanl.gov/abs/0706.0398}{{\tt arXiv:0706.0398}}].

\bibitem{Chamseddine:2007hz}
A.~H. Chamseddine and A.~Connes, {\it {Why the Standard Model}},  {\em J. Geom.
  Phys.} {\bf 58} (2008) 38--47, [\href{http://xxx.lanl.gov/abs/0706.3688}{{\tt
  arXiv:0706.3688}}].

\bibitem{Grosse:2008xr}
H.~Grosse, H.~Steinacker and M.~Wohlgenannt, {\it {Emergent Gravity, Matrix
  Models and UV/IR Mixing}},  {\em JHEP} {\bf 04} (2008) 023,
  [\href{http://xxx.lanl.gov/abs/0802.0973}{{\tt arXiv:0802.0973}}].

\bibitem{Bietenholz:2002ch}
W.~Bietenholz, F.~Hofheinz and J.~Nishimura, {\it {The Renormalizability of
  2-D Yang-Mills theory on a noncommutative geometry}},  {\em JHEP} {\bf 0209}
  (2002) 009, [\href{http://xxx.lanl.gov/abs/hep-th/0203151}{{\tt
  hep-th/0203151}}].

\bibitem{Azuma:2004zq}
T.~Azuma, S.~Bal, K.~Nagao and J.~Nishimura, {\it {Nonperturbative studies of
  fuzzy spheres in a matrix model with the Chern-Simons term}},  {\em JHEP}
  {\bf 05} (2004) 005, [\href{http://xxx.lanl.gov/abs/hep-th/0401038}{{\tt
  hep-th/0401038}}].

\bibitem{CastroVillarreal:2004vh}
P.~Castro-Villarreal, R.~Delgadillo-Blando and B.~Ydri, {\it {A
  Gauge-invariant UV-IR mixing and the corresponding phase transition for U(1)
  fields on the fuzzy sphere}},  {\em Nucl. Phys.} {\bf B704} (2005) 111--153,
  [\href{http://xxx.lanl.gov/abs/hep-th/0405201}{{\tt hep-th/0405201}}].

\bibitem{O'Connor:2006wv}
D.~O'Connor and B.~Ydri, {\it {Monte Carlo Simulation of a NC Gauge Theory on
  The Fuzzy Sphere}},  {\em JHEP} {\bf 11} (2006) 016,
  [\href{http://xxx.lanl.gov/abs/hep-lat/0606013}{{\tt hep-lat/0606013}}].

\bibitem{Bietenholz:2006cz}
W.~Bietenholz, J.~Nishimura, Y.~Susaki and J.~Volkholz, {\it {A
  Non-perturbative study of 4-D U(1) non-commutative gauge theory: The Fate of
  one-loop instability}},  {\em JHEP} {\bf 0610} (2006) 042,
  [\href{http://xxx.lanl.gov/abs/hep-th/0608072}{{\tt hep-th/0608072}}].

\bibitem{DelgadilloBlando:2007vx}
R.~Delgadillo-Blando, D.~O'Connor and B.~Ydri, {\it {Geometry in Transition: A
  Model of Emergent Geometry}},  {\em Phys. Rev. Lett.} {\bf 100} (2008)
  201601, [\href{http://xxx.lanl.gov/abs/0712.3011}{{\tt arXiv:0712.3011}}].

\bibitem{Azeyanagi:2008bk}
T.~Azeyanagi, M.~Hanada and T.~Hirata, {\it {On Matrix Model Formulations of
  Noncommutative Yang-Mills Theories}},  {\em Phys. Rev.} {\bf D78} (2008)
  105017, [\href{http://xxx.lanl.gov/abs/0806.3252}{{\tt arXiv:0806.3252}}].

\bibitem{Spisso:2011wy}
B.~Spisso and R.~Wulkenhaar, {\it {A numerical approach to harmonic
  non-commutative spectral field theory}},  {\em Int. J. Mod. Phys.} {\bf A27}
  (2012) 1250075, [\href{http://xxx.lanl.gov/abs/1111.3050}{{\tt
  arXiv:1111.3050}}].

\bibitem{O'Connor:2013rla}
D.~O'Connor, B.~P. Dolan and M.~Vachovski, {\it {Critical Behaviour of the
  Fuzzy Sphere}},  {\em JHEP} {\bf 12} (2013) 085,
  [\href{http://xxx.lanl.gov/abs/1308.6512}{{\tt arXiv:1308.6512}}].

\bibitem{Hayakawa:1999yt}
M.~Hayakawa, {\it {Perturbative analysis on infrared aspects of noncommutative
  QED on {$R^4$}}},  {\em Phys. Lett.} {\bf B478} (2000) 394--400,
  [\href{http://xxx.lanl.gov/abs/hep-th/9912094}{{\tt hep-th/9912094}}].

\bibitem{Martin:2000bk}
C.~P. Mart{\'{\i}}n and F.~Ruiz~Ruiz, {\it {Paramagnetic dominance, the sign of
  the beta function and UV / IR mixing in noncommutative U(1)}},  {\em Nucl.
  Phys.} {\bf B597} (2001) 197--227,
  [\href{http://xxx.lanl.gov/abs/hep-th/0007131}{{\tt hep-th/0007131}}].

\bibitem{VanRaamsdonk:2001jd}
M.~Van~Raamsdonk, {\it {The Meaning of infrared singularities in noncommutative
  gauge theories}},  {\em JHEP} {\bf 11} (2001) 006,
  [\href{http://xxx.lanl.gov/abs/hep-th/0110093}{{\tt hep-th/0110093}}].

\bibitem{Bluemer:2009zf}
J.~Bl{\"u}mer, R.~Engel and J.~R. H{\"o}randel, {\it {Cosmic Rays from the
  Knee to the Highest Energies}},  {\em Prog. Part. Nucl. Phys.} {\bf 63}
  (2009) 293--338, [\href{http://xxx.lanl.gov/abs/0904.0725}{{\tt
  arXiv:0904.0725}}].

\bibitem{Szabo:2009tn}
R.~J. Szabo, {\it {Quantum Gravity, Field Theory and Signatures of
  Noncommutative Spacetime}},  {\em Gen. Rel. Grav.} {\bf 42} (2010) 1--29,
  [\href{http://xxx.lanl.gov/abs/0906.2913}{{\tt arXiv:0906.2913}}].

\bibitem{Banks:1996vh}
T.~Banks, W.~Fischler, S.~H. Shenker and L.~Susskind, {\it {M theory as a
  matrix model: A Conjecture}},  {\em Phys. Rev.} {\bf D55} (1997) 5112--5128,
  [\href{http://xxx.lanl.gov/abs/hep-th/9610043}{{\tt hep-th/9610043}}].

\bibitem{Witten:1995ex}
E.~Witten, {\it {String theory dynamics in various dimensions}},  {\em Nucl.
  Phys.} {\bf B443} (1995) 85--126,
  [\href{http://xxx.lanl.gov/abs/hep-th/9503124}{{\tt hep-th/9503124}}].

\bibitem{Ishiki:2008te}
G.~Ishiki, S.-W. Kim, J.~Nishimura and A.~Tsuchiya, {\it {Deconfinement phase
  transition in {$\mathcal{N}=4$} super Yang-Mills theory on {$R \times S^3$}
  from supersymmetric matrix quantum mechanics}},  {\em Phys. Rev. Lett.} {\bf
  102} (2009) 111601, [\href{http://xxx.lanl.gov/abs/0810.2884}{{\tt
  arXiv:0810.2884}}].

\bibitem{Ambjorn:2000bf}
J.~Ambj{\o}rn, K.~N. Anagnostopoulos, W.~Bietenholz, T.~Hotta, and
  J.~Nishimura, {\it {Large N dynamics of dimensionally reduced 4-D SU(N)
  superYang-Mills theory}},  {\em JHEP} {\bf 07} (2000) 013,
  [\href{http://xxx.lanl.gov/abs/hep-th/0003208}{{\tt hep-th/0003208}}].

\bibitem{Ambjorn:2000dx}
J.~Ambj{\o}rn, K.~N. Anagnostopoulos, W.~Bietenholz, T.~Hotta, and
  J.~Nishimura, {\it {Monte Carlo studies of the IIB matrix model at large N}},
   {\em JHEP} {\bf 07} (2000) 011,
  [\href{http://xxx.lanl.gov/abs/hep-th/0005147}{{\tt hep-th/0005147}}].

\bibitem{Ambjorn:2001xs}
J.~Ambj{\o}rn, K.~N. Anagnostopoulos, W.~Bietenholz, F.~Hofheinz, and
  J.~Nishimura, {\it {On the spontaneous breakdown of Lorentz symmetry in
  matrix models of superstrings}},  {\em Phys. Rev.} {\bf D65} (2002) 086001,
  [\href{http://xxx.lanl.gov/abs/hep-th/0104260}{{\tt hep-th/0104260}}].

\bibitem{Anagnostopoulos:2005cy}
K.~N. Anagnostopoulos, T.~Azuma, K.~Nagao and J.~Nishimura, {\it {Impact of
  supersymmetry on the nonperturbative dynamics of fuzzy spheres}},  {\em JHEP}
  {\bf 09} (2005) 046, [\href{http://xxx.lanl.gov/abs/hep-th/0506062}{{\tt
  hep-th/0506062}}].

\bibitem{Hanada:2007ti}
M.~Hanada, J.~Nishimura and S.~Takeuchi, {\it {Non-lattice simulation for
  supersymmetric gauge theories in one dimension}},  {\em Phys. Rev. Lett.}
  {\bf 99} (2007) 161602, [\href{http://xxx.lanl.gov/abs/0706.1647}{{\tt
  arXiv:0706.1647}}].

\bibitem{Anagnostopoulos:2007fw}
K.~N. Anagnostopoulos, M.~Hanada, J.~Nishimura and S.~Takeuchi, {\it {Monte
  Carlo studies of supersymmetric matrix quantum mechanics with sixteen
  supercharges at finite temperature}},  {\em Phys. Rev. Lett.} {\bf 100}
  (2008) 021601, [\href{http://xxx.lanl.gov/abs/0707.4454}{{\tt
  arXiv:0707.4454}}].

\bibitem{Kawahara:2007eu}
N.~Kawahara, J.~Nishimura and A.~Yamaguchi, {\it {Monte Carlo approach to
  nonperturbative strings - Demonstration in noncritical string theory}},  {\em
  JHEP} {\bf 06} (2007) 076,
  [\href{http://xxx.lanl.gov/abs/hep-th/0703209}{{\tt hep-th/0703209}}].

\bibitem{Catterall:2008yz}
S.~Catterall and T.~Wiseman, {\it {Black hole thermodynamics from simulations
  of lattice Yang-Mills theory}},  {\em Phys. Rev.} {\bf D78} (2008) 041502,
  [\href{http://xxx.lanl.gov/abs/0803.4273}{{\tt arXiv:0803.4273}}].

\bibitem{Hanada:2009ne}
M.~Hanada, J.~Nishimura, Y.~Sekino and T.~Yoneya, {\it {Monte Carlo studies of
  Matrix theory correlation functions}},  {\em Phys. Rev. Lett.} {\bf 104}
  (2010) 151601, [\href{http://xxx.lanl.gov/abs/0911.1623}{{\tt
  arXiv:0911.1623}}].

\bibitem{Hanada:2012si}
M.~Hanada, M.~Honda, Y.~Honma, J.~Nishimura, S.~Shiba and Y.~Yoshida, {\it
  {Numerical studies of the ABJM theory for arbitrary N at arbitrary coupling
  constant}},  {\em JHEP} {\bf 05} (2012) 121,
  [\href{http://xxx.lanl.gov/abs/1202.5300}{{\tt arXiv:1202.5300}}].

\bibitem{Anagnostopoulos:2013xga}
K.~N. Anagnostopoulos, T.~Azuma and J.~Nishimura, {\it {Monte Carlo studies of
  the spontaneous rotational symmetry breaking in dimensionally reduced super
  Yang-Mills models}},  {\em JHEP} {\bf 11} (2013) 009,
  [\href{http://xxx.lanl.gov/abs/1306.6135}{{\tt arXiv:1306.6135}}].

\bibitem{Hanada:2013rga}
M.~Hanada, Y.~Hyakutake, G.~Ishiki and J.~Nishimura, {\it {Holographic
  description of quantum black hole on a computer}},  {\em Science} {\bf 344}
  (2014) 882--885, [\href{http://xxx.lanl.gov/abs/1311.5607}{{\tt
  arXiv:1311.5607}}].

\bibitem{Filev:2015hia}
V.~G. Filev and D.~O'Connor, {\it {The BFSS model on the lattice}},
  \href{http://xxx.lanl.gov/abs/1506.01366}{{\tt arXiv:1506.01366}}.

\bibitem{Filev:2015cmz}
V.~G. Filev and D.~O'Connor, {\it {A Computer Test of Holographic Flavour
  Dynamics}},  \href{http://xxx.lanl.gov/abs/1512.02536}{{\tt arXiv:1512.02536}}.

\bibitem{Kawahara:2007ib}
N.~Kawahara, J.~Nishimura and S.~Takeuchi, {\it {High temperature expansion in
  supersymmetric matrix quantum mechanics}},  {\em JHEP} {\bf 12} (2007) 103,
  [\href{http://xxx.lanl.gov/abs/0710.2188}{{\tt arXiv:0710.2188}}].

\bibitem{Klebanov:1996un}
I.~R. Klebanov and A.~A. Tseytlin, {\it {Entropy of near extremal black
  p-branes}},  {\em Nucl. Phys.} {\bf B475} (1996) 164--178,
  [\href{http://xxx.lanl.gov/abs/hep-th/9604089}{{\tt hep-th/9604089}}].

\bibitem{Hanada:2008gy}
M.~Hanada, A.~Miwa, J.~Nishimura and S.~Takeuchi, {\it {Schwarzschild radius
  from Monte Carlo calculation of the Wilson loop in supersymmetric matrix
  quantum mechanics}},  {\em Phys. Rev. Lett.} {\bf 102} (2009) 181602,
  [\href{http://xxx.lanl.gov/abs/0811.2081}{{\tt arXiv:0811.2081}}].

\bibitem{Rey:1998ik}
S.-J. Rey and J.-T. Yee, {\it {Macroscopic strings as heavy quarks in large N
  gauge theory and anti-de Sitter supergravity}},  {\em Eur. Phys. J.} {\bf
  C22} (2001) 379--394, [\href{http://xxx.lanl.gov/abs/hep-th/9803001}{{\tt
  hep-th/9803001}}].

\bibitem{Anagnostopoulos:2001yb}
K.~N. Anagnostopoulos and J.~Nishimura, {\it {New approach to the
  complex-action problem and its application to a nonperturbative study of
  superstring theory}},  {\em Phys.Rev.} {\bf D66} (2002) 106008,
  [\href{http://xxx.lanl.gov/abs/hep-th/0108041}{{\tt hep-th/0108041}}].

\bibitem{Anagnostopoulos:2010ux}
K.~N. Anagnostopoulos, T.~Azuma and J.~Nishimura, {\it {A General approach to
  the sign problem: The Factorization method with multiple observables}},  {\em
  Phys. Rev.} {\bf D83} (2011) 054504,
  [\href{http://xxx.lanl.gov/abs/1009.4504}{{\tt arXiv:1009.4504}}].

\bibitem{deForcrand:2010ys}
P.~de~Forcrand, {\it {Simulating QCD at finite density}},  {\em PoS} {\bf
  Lattice 2009} (2009) 010, [\href{http://xxx.lanl.gov/abs/1005.0539}{{\tt
  arXiv:1005.0539}}].

\bibitem{Loh:1990sp}
E.~Y. Loh, J.~E. Gubernatis, R.~T. Scalettar, S.~R. White, D.~J. Scalapino, and
  R.~L. Sugar, {\it {Sign problem in the numerical simulation of many-electron
  systems}},  {\em Phys.Rev.} {\bf B41} (1990) 9301.

\end{thebibliography}\endgroup

\end{document}